\documentclass[preprint,showpacs,preprintnumbers,amsmath,amssymb,nofootinbib]{revtex4}
\usepackage{bbm}
\usepackage{amsfonts}
\usepackage{booktabs}
\usepackage{mathrsfs}
\usepackage{epsfig}
\usepackage{graphicx}
\usepackage{dcolumn}
\usepackage{bm}
\usepackage{amsmath}
\usepackage{slashed}

\let\jnfont=\rm
\def\NPB#1,{{\jnfont Nucl.\ Phys.\ B }{\bf #1},}
\def\PLB#1,{{\jnfont Phys.\ Lett.\ B }{\bf #1},}
\def\EPJC#1,{{\jnfont Eur.\ Phys.\ Jour.\ C }{\bf #1},}
\def\PRD#1,{{\jnfont Phys.\ Rev.\ D }{\bf #1},}
\def\PRL#1,{{\jnfont Phys.\ Rev.\ Lett.\ }{\bf #1},}
\def\MPLA#1,{{\jnfont Mod.\ Phys.\ Lett.\ A }{\bf #1},}
\def\JPG#1,{{\jnfont J.\ Phys.\ G}{\bf #1},}
\def\CTP#1,{{\jnfont Commun.\ Theor.\ Phys.\ }{\bf #1},}
\def\ZPC#1,{{\jnfont Z.\ Phys.\ C }{\bf #1},}
\def\JHEP#1,{{\jnfont JHEP \ }{\bf #1},}
\def\Rv{\not{\hbox{\kern-1pt $R$}}}
\def\p{\not{\hbox{\kern-3pt $p$}}}

\begin{document}
\preprint{\parbox{1.2in}{\noindent arXiv:}}

\title{Probing light bino and higgsinos at the LHC}

\author{ Chengcheng Han
        \\~ \vspace*{-0.3cm} }
\affiliation{ Asia Pacific Center for Theoretical Physics,
             San 31, Hyoja-dong, Nam-gu,
             Pohang 790-784, Republic of Korea \\
             \vspace*{1.5cm}}

\begin{abstract}
Motivated by the naturalness, we study a simplified MSSM scenario where only the bino-like LSP and higgsino-like NLSP are light. We first scan the parameter space of this scenario, considering the constraints from the Higgs mass, flavor physics, electroweak precision measurements and dark matter experiments. Then in the allowed parameter space, we perform a Monte Carlo simulation for the $\tilde{\chi}^\pm_1 \tilde{\chi}^0_{2,3}$ production followed by $\tilde{\chi}^\pm_1 \to W^\pm \tilde{\chi}^0_1$ and $\tilde{\chi}^0_{2,3} \to Z\tilde{\chi}^0_1$. By examining the presently available trilepton bounds on the wino-like chargino/neutralino, we find that only a narrow region $40\,\rm{GeV} \lesssim m_{\tilde{\chi}^0_1} \lesssim 50\,\rm{GeV}$ and $160\,\rm{GeV} \lesssim m_{\tilde{\chi}^0_{2,3}} \lesssim 170\,\rm {GeV}$ on the plane of $m_{\tilde{\chi}^0_1}-m_{\tilde{\chi}^0_{2,3}}$ can be excluded. Finally, we explore the potential of trilepton signature in probing such a scenario at 14 TeV LHC and find that the region with $40\,\rm{GeV} \lesssim m_{\tilde{\chi}^0_1} \lesssim 60\,\rm {GeV}$ and $160 \rm {GeV}\,\lesssim m_{\tilde{\chi}^0_{2,3}} \lesssim 300\,\rm{GeV}$ can be covered at $3\sigma$ level with luminosity ${\cal L}=300$ fb$^{-1}$.
\end{abstract}

\maketitle
\section{introduction}
The ATLAS and CMS have observed the Higgs boson \cite{ATLAS-CMS} and up to now the measurements of
Higgs properties consist with the standard model (SM) predictions. However, the SM suffers from the so-called
naturalness problem \cite{finetuning} which inspires theorists to propose various new physics models.
Among these new physics models the natural SUSY \cite{nsusy1,nsusy2,nsusy3,nsusy4,nsusy5} satisfied
the naturalness criterion perfectly. It needs a light stop sector and a weak scale higgsino mass
$\mu \lesssim O(300\,\text{GeV})$. The stop sector in the natural SUSY has been discussed extensively \cite{stop,nsusy-stop,hanz,drees}.
The weak scale higgsino in the natural SUSY results in the existence of at least two light neutralinos and a pair of charginos at the weak scale.
 If the lighter one of the two neutralinos is the dark matter, the dark matter relic density  would be far below the observed one because a higgsino like
dark matter usually has large annihilation cross section to the SM particles.  However if the bino is the LSP, then the dark matter is bino like and
has some higgsino component, the dark matter relic density could much easier to be satisfied.  So searching for such kinds of electroweakinos
would directly probe the dark matter sector and the naturalness of SUSY.

Generally, the search strategies depends on the spectrum of electroweakinos. If the electroweakinos are highly degenerate at low energy, they could be probed by the mono-jet, mono-photon or mono-$Z$ in the future experiments \cite{monojet,carpenter,liutao}. One such case is, at weak scale, only the higgsinos are light.
In this case, the mono-jet signal can search higgsinos to 150 GeV at 2$\sigma$ level at 14TeV LHC
with luminosity $3000~\rm{fb}^{-1}$.

If the mass splitting between the electroweakinos is moderate, they can be probed through multi-soft
leptons \cite{soft-lepton}. Recently, some authors also proposed a new channel
$\ell^+ \ell^- +\gamma +\slashed E_T $ to probe the region with a small splitting between the higgsinos
and bino \cite{bino-higgsino}.
The photon in the final state comes from the
 $\chi^0_2$ decaying into $\chi^0_1$ plus a photon, and the two leptons come from the other neutralino
decaying into LSP via a virtual $Z$ boson. When the splitting between the two neutralinos is small,
the branching ratio  of $\chi_2^0\rightarrow \gamma \chi_0^1$ is considerable. So, another signal
$j+\ell+ \gamma + \slashed {E}_T $ from neutralino/chargino pair production may be also accessible
at the future LHC \cite{new}.

If the electroweakinos have a large mass splitting, the multi-leptons final state
from chargino/neutralino pair production has the highest sensitivity \cite{higgsino-3l,hantao}.
The ATLAS and CMS collaborations performed such a study and gave the mass limits
$m_{\tilde{\chi}^\pm_1,~\tilde{\chi}^0_2} >345$ GeV (ATLAS) \cite{atlas-3l} and
$m_{\tilde{\chi}^\pm_1,~\tilde{\chi}^0_2} >270$ GeV (CMS) \cite{cms-3l} assuming
the chargino/neutralino decays via intermediate gauge bosons and the LSP is almost massless.
 But this limit depends on the component of chargino/neutralino.
In the experiment, a wino NLSP and bino LSP are assumed and thus the pair produced
chargino/neutralino are wino-like. The limit would be changed if the NLSP is higgsino.
One reason is that the cross section of chargino/neutralino production in the wino NLSP case
is larger than the one in the higgsino case. The other reason is that in the realistic spectrum,
the decay branching ratio of $\chi_2^0\rightarrow\chi^0_1 + Z$  is not 100\% since $\chi_2^0$ could decay into $\chi^0_1 + h$.

In the present work, we focuses on the LSP bino and NLSP higgsino case. We also require the wino decoupled.
 Such kinds of spectrums can be realized in the non-universal gaugino masses models.
We reinterpreted the experiments results in the realistic spectrum, and give the prospect of detecting this signal
in the future LHC experiments.

The rest of this paper is organized as follows:
In Sec. II, we scan the parameter space, and the properties of the surviving parameter space are investigated.
In Sec. III, we reinterpret the experimental limits on the parameter space.
In Sec. IV, the prospect of detecting this signal in our surviving space are studied.

\section{The property of surviving parameter space }

In this section, we scan the parameter space in the frame of natural SUSY with a light bino. Some survived samples are shown, which are susceptible to the 3$l$ experiment and possible closed by future direct search results.

The parameter space are scanned in the following region:
\begin{eqnarray}
    1~\text{GeV} <M_1< 100~\text{GeV},\quad 100~\text{GeV}<\mu < 300~\text{GeV},\quad 3 < \tan\beta < 60 ,
\end{eqnarray}
where the lower bound of $\mu$ avoids the chargino search limit and upper bound satisfies the naturalness requirement.
Other parameters, except for the stop sector, are fix at 2 TeV. The stop sector are scanned in this region,
\begin{eqnarray}
700 ~\textrm{GeV} <(M_{\tilde{Q}_3},M_{\tilde{t}_R})< 2 ~\textrm{TeV},
~~-3 ~\textrm{TeV}< A_t < 3~\textrm{TeV}
\end{eqnarray}
where the lower bound avoids the direct stop search limit and the upper bound keeps the naturalness of the SUSY.

The following constraints are considered in the scan:
\begin{itemize}
  \item[(1)]  The SM-like Higgs mass is required to within the range of 123--127 GeV.
    We use \textsf{FeynHiggs2.8.9} \cite{feynhiggs} to calculate the Higgs mass, and impose the experimental
    constraints from LEP, Tevatron and LHC by \textsf{HiggsBounds-3.8.0} \cite{higgsbounds}.
  \item[(2)] We require our samples to satisfy various B-physics bounds at 2$\sigma$ level.
    We use \textsf{SuperIso v3.3} \cite{superiso} to implement the constraints, including
    $B\rightarrow X_s\gamma$ and the latest measurements of $B_s\rightarrow \mu^+\mu^-$,
    $B_d\rightarrow X_s\mu^+\mu^-$ and $ B^+\rightarrow \tau^+\nu$.
  \item[(3)] The SUSY prediction of the precision electroweak observable, such as
    $\rho_l$, $\sin^2 \theta_{\rm eff}^l$, $m_W$ and $R_b$ \cite{rb}, are required to be
    within the $2\sigma$ ranges of the experimental values.
  \item[(4)] A light bino in the natural SUSY would mix with the higgsino, which induces three neutralinos and a pair of charginos.
    The lightest neutralino acts as the dark matter candidate.
    So the relic abundance and the direct search of the dark matter set limit on the parameter space.
    Here we require the thermal relic density of the lightest neutralino (as the dark matter candidate)
    is under the 2$\sigma$ upper limit of the Planck value \cite{planck}.  We use the code \textsf{MicrOmega v2.4} \cite{micromega}
    to calculate the relic abundance and DM-nucleon scattering.
\end{itemize}

\begin{figure}[htbp]
\includegraphics[width=6.0in,height=3in]{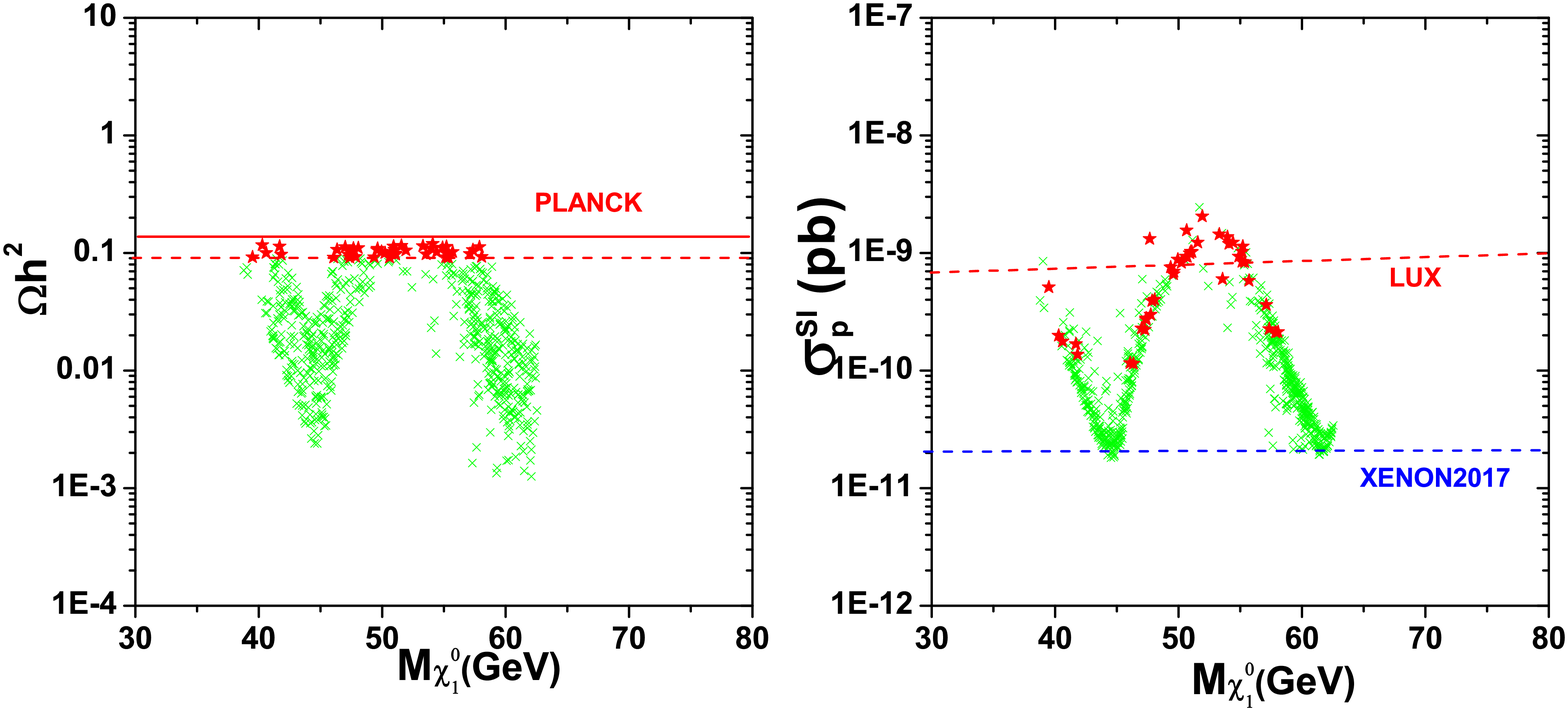}
\caption{ Dark matter properties in the surviving space. All points satisfy 2$\sigma$ upper limit of PLANCK. The red pentagrams locate within the 2$\sigma$ range of PLANCK, $0.091 < \Omega h^2 < 0.138$, where a 10\% theoretical uncertainty is included. The DM-nucleon scattering cross section has been scaled by a factor $\Omega h^2/\Omega h^2(PLANCK)$. }
\label{fig1}
\end{figure}

Figure \ref{fig1} shows the dark matter properties in the surviving space.
The left panel scatters the relic abundance of the surviving samples.
We can see that there are two regions left both of which have a bino like dark matter, requiring the 2$\sigma$ upper limit of the Planck value \cite{planck}.
The first region corresponding a lighter dark matter mass lying in $35\sim 50\,\text{GeV}$ near half of the Z boson mass.
In the other region, the dark matter mass lies in $50-65$ GeV near half of the Higgs mass. It is easy to explain it.
Usually, a bino LSP has a small annihilation cross section and thus it is easy to get a large relic density.
 If there is a particle in the s-channel, the annihilation cross section can be raised through resonance enhancement and thus the relic abundance are reduced.
It requires that the dark matter mass is around half of the boson mass. In our spectrum, only the $Z$ boson or the Higgs boson can play this role.
We should note there are points satisfying the 2$\sigma$ range of PLANCK. These points could get the correct relic abundance due to
joint effect of the resonance and the mixing between bino and higgsino \cite{well-temper}.

The right panel shows the limit of dark matter direct search.
It tells that the $Z/h$ resonance region easily get rid of the LUX \cite{LUX} constraint, due to the moderate spitting between bino and higgsino. The future XENON-1T (2017) will exclude large parameter space of this region, but a small fraction can still survived when dark matter mass is very close to half the $Z$ mass or Higgs mass. It should be noted the some points within the 2$\sigma$ range of PLANCK still survived the LUX search. But they are possibly covered by future dark matter Direct search  XENON-1T (2017).

So, in the following, we concentrate on the points which are still survived under limits from the dark matter experiments.
Although in this region, a correct relic abundance could be derived by elaborately tuning the neutralino mass, we just take the dark matter relic abundance as an upper limit here.

\section{Direct search limits on the parameter space}

At LHC, the ATLAS and CMS collaborations have separately preform the $3l$ searches \cite{atlas-3l,cms-3l}. They aims at the $\chi^0_2 \chi^{\pm}_1$ pair production following by decays $\chi_2^0\rightarrow\chi^0_1 + Z$, $\chi^\pm_1\rightarrow\chi_1^0 + W^\pm$ (the $W$ and $Z$ can be virtual), and then the $W$/$Z$ decays producing 3 leptons in the final state. In this paper, we use the ATLAS experiment result to constrain our parameter space.

The ATLAS experiment\cite{atlas-3l2} defines six signal regions aiming at $Z$-depleted region and $Z$-enriched region. Table \ref{tab1} shows the selection requirements of these six signal regions. We can see that SRnoZ{(a,b,c)} concentrate on the $Z$-depleted case where the invariant mass of the SFOS lepton pair departs the $Z$-boson mass. Conversely, in the other three regions, a $Z$ boson is required in the mediate state.

Through the analysis, the experiments give an exclusion region on the $\chi^0_2-\chi^0_1$ plane  and the exclusion limit can reach 320 GeV \footnote{We note that in the latest ATLAS $3l$ results this limit reaches 345 GeV. We carefully check the difference between the old one and the latest one, and find that the cut efficiency is not improved significantly and the latest result is also hard to implement because twenty signal regions are defined.}.
It should be noted that in the experiments a wino like NLSP and bino LSP are assumed and the decay branching ratio of $\chi_2^0\rightarrow\chi^0_1 + Z$ and $\chi^\pm_1\rightarrow\chi_1^0 + W^\pm$ are set to be 100\%.

In the higgsino NLSP case, not only the $\chi^0_2$, bust also the $\chi^0_3$ contributes to the signals.
Even so, the total cross section is still about half of the one in the wino case.
So we should carefully implement the $3l$ experiments on our parameter space; in the present work, we use the Monte Carlo simulation.
\textsf{MadGraph5} \cite{mad5} are adopted to generate events, the parton shower is carried out by \textsf{PYTHIA} \cite{pythia}, and \textsf{CheckMATE1.1.4} \cite{Checkmate} are used to simulate the $3l$ experiments events.
Finally, we combine the simulation results from the six signal regions and derive the final exclusion limit.
At the beginning, we checked the reliability of our simulation with the benchmark points provided by ATLAS paper; we found them well consistent.

Figure 2 shows our limit assuming a 100\% branching ratio of $\chi_2^0\rightarrow\chi^0_1 + Z$ and $\chi_3^0\rightarrow\chi^0_1 + Z$.
We can see the limit can reach utmost 250 GeV when the LSP is near to 0 GeV, comparing to the limit 320 GeV in the wino case. It also tells when the LSP becomes heavy, the limit reduces rapidly.
This figure is also consistent with similar figures provided by other authors\cite{higgsino-3l}.
We should note that there is a part where the limit is not effective. It locates around $M_{\chi^0_2} \sim$ $140-160$ GeV and $M_{\chi^0_1} \sim$ $40-60$ GeV.
It can be explained as follows. When the splitting of $\chi^0_2$ and $\chi^0_1$ is just larger than the $Z$ mass, this kinematics is very similar to the one of $WZ$ background and thus its backgrounds are relatively large. Then this region has a small probing efficiency.

\begin{table}[th] \centering\caption{The selection requirements for the six signal regions. \label{tab1}}
\begin{tabular}{|c|c|c|c|c|c|c|c|}
\hline  Selection                 & SRnoZa         & SRnoZb      & SRnoZc   & SRZa  & SRZb & SRZc \\
\hline  $m_{SFOS}$ [GeV]                         &$<$ 60 	&60-81.2 	&$<$81.2 or $>$ 101.2 	&81.2-101.2 	 &81.2-101.2 	&81.2-101.2  \\
\hline  $E_T^{miss}$ [GeV]                   &$ > $50 	& $>$75 	&$>$75 	&75-120 	&75-120 	&$>$120  \\
\hline  $m_T$ [GeV]                &$-$ 	&$-$	&$>$110 	&$<$110 	&$>$110 	&$>$110  \\
\hline  $p_T$ $3^{rd}~l$ [GeV]                &$>$10 	&$>$10 	&$>$30 	&$>$10 	&$>$10 	&$>$10  \\
\hline  SR veto           & SRnoZc   & SRnoZc & $-$ & $-$ & $-$ & $-$ \\
\hline
\end{tabular}
\label{tab1}
\end{table}

\begin{figure}[htbp]
\includegraphics[width=4.0in,height=3in]{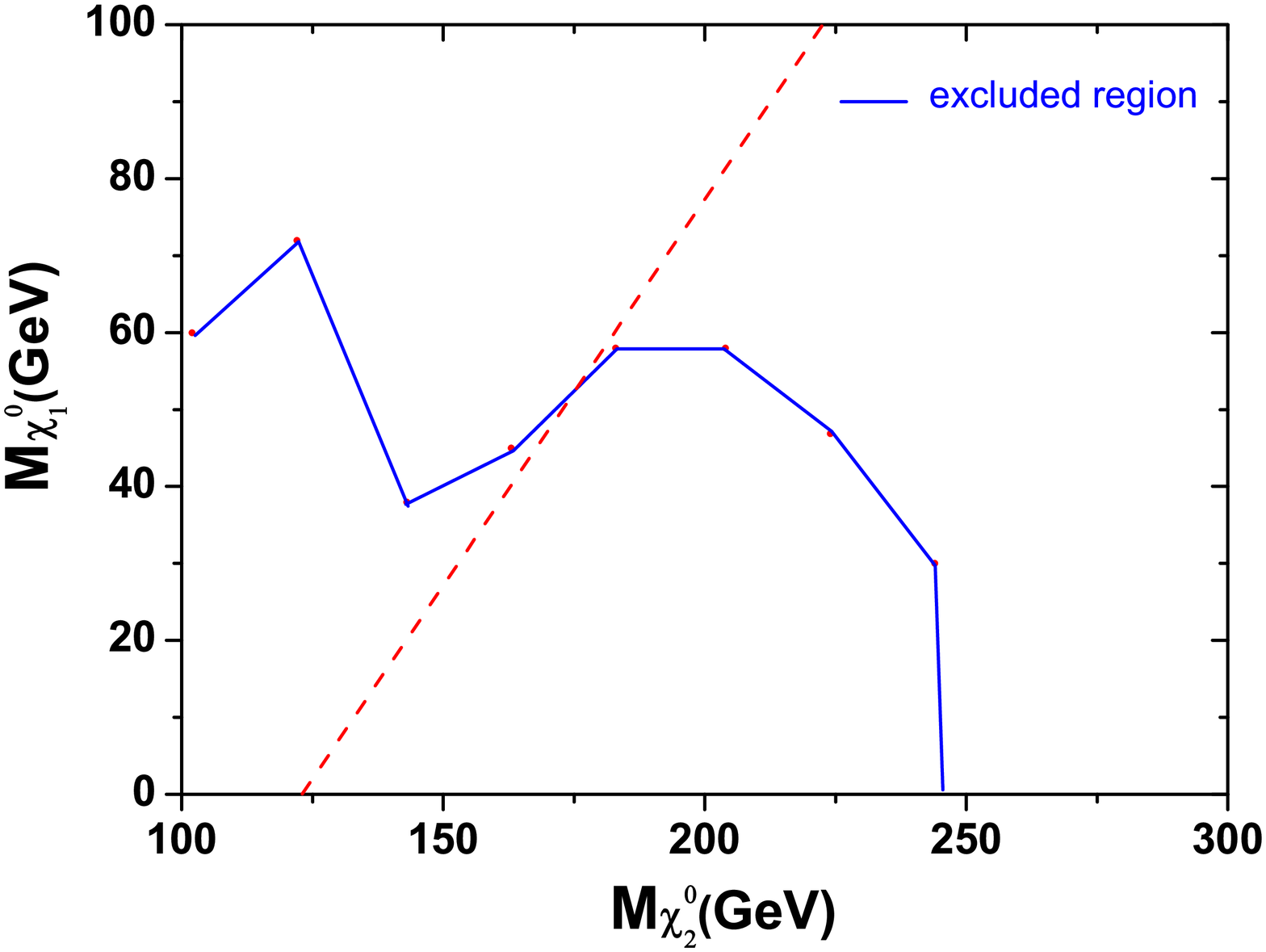}
\caption{The $3l$ exclusion limit on the $\chi^0_2-\chi^0_1$ plane with higgsino being the NLSP and a 100\% branching ratio of $\chi_{2,3}^0\rightarrow\chi^0_1 + Z$ is assumed. The red dashed line is the dividing line $m_{\chi_2^0}=m_{\chi_1^0}+m_{h}$. On the right part of this line, the decay channel of $\chi^0_{2}\rightarrow\chi^0_1 h$ opens.}
\label{fig2}
\end{figure}

To impose the $3l$ constraint on our samples, it should also survey the decay branching ratio of $\chi^0_{2,3}$.
Fig. 3 presents the decays of $\chi^0_{2,3}$ ($\chi^0_{2,3}\rightarrow\chi^0_1 Z$ including off-shell Z). When $\chi^0_1 h $ channel does not open, the $\chi^0_{2,3}\rightarrow\chi^0_1 Z$ dominates. Otherwise, we can see two distinct regions for the decays of $\chi^0_{2,3}$, the $h$-enriched region and the $h$-depressed region. In the $h$-enriched region, the branching ratio of $\chi^0_{2,3}\rightarrow\chi^0_1 h$ can reach as much as $75\%$, whereas in the $h$-depressed region, this branching ratio would be less than 25\%.

Note that $\chi^0_{2}\rightarrow\chi^0_1 h$ and $\chi^0_{3}\rightarrow\chi^0_1 h$ can not be enriched at the same time, which can be inferred from the third panel of Fig. 3. The reason is illustrated clearly in paper\cite{hantao,jung}. It also should point out that the $3l$ probing ability is relevant to the $Br(\chi^0_{2}\rightarrow\chi^0_1 Z) +Br(\chi^0_{3}\rightarrow\chi^0_1 Z)$.

\begin{figure}[htbp]
\includegraphics[width=7.0in,height=3in]{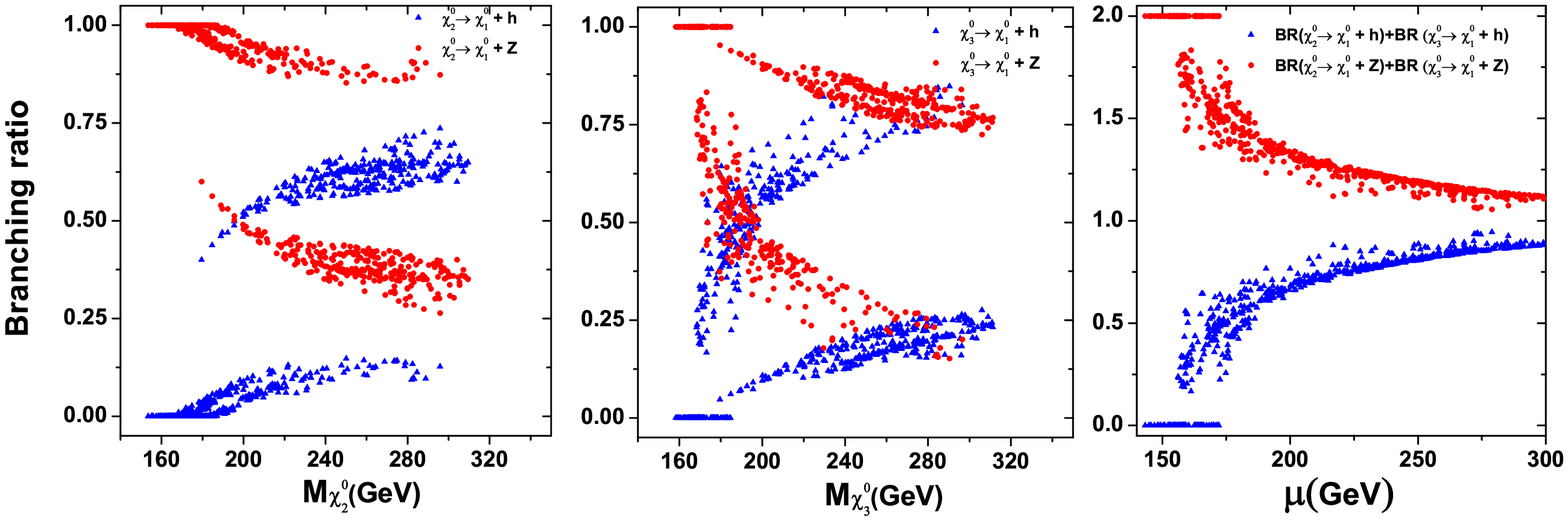}
\caption{The decay branching ratio of the neutralinos.}
\label{fig3}
\end{figure}

\begin{figure}[htbp]
\includegraphics[width=4.0in,height=3in]{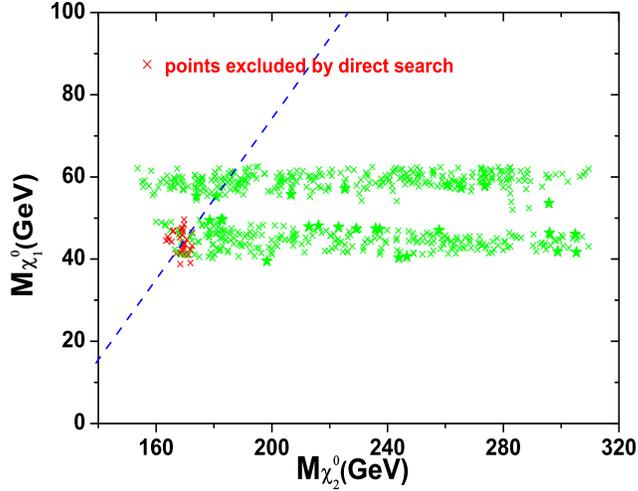}
\caption{The $3l$ exclusion results considering branching ratio effect. The blue dashed line is the dividing line $m_{\chi_2^0}=m_{\chi_1^0}+m_{h}$. On the right part of this line, the decay channel of $\chi^0_{2}\rightarrow\chi^0_1 h$ opens. The pentagrams on the plot satisfying the 2$\sigma$ range of PLANCK.}
\label{fig4}
\end{figure}

Figure 4 presents the limit at 8 TeV LHC on the parameter space considering the branching ratio effect.
It is found that only a very tiny region can be excluded in this scenario and most regions survived. The blue line
represents the threshold $\chi^0_{2}\rightarrow\chi^0_1 h$. Only the points near the bottom of the line have been
excluded. The points which survives lying the left part of the line escape from the experimental limits
largely due to a suppression on the the kinematics cut. And the reason for the points surviving on the right part
of the line is the $\chi^0_{2}\rightarrow\chi^0_1 h$ channel opens and the reduce of $\chi^0_{2,3}\chi^\pm_1$
production rate.
Note that even the $\chi^0_{2}\rightarrow\chi^0_1 h$ channel opens, there are still points excluded.
The reason is that when the $\chi^0_{2}\rightarrow\chi^0_1 h$ is just open, the branching ratio of
$\chi^0_{2,3}\rightarrow\chi^0_1 h$ is still very small, which can be seen from the right panel of Fig. 3.
We note the points satisfying the relic density are still relatively safe.

In addition, the mass of $\chi^0_2$ in our surviving samples is larger than 150 GeV.
It is largely due to the Z invisible decay and Higgs invisible decay limit because a fraction of Higgsino component of
dark matter would sizably affect the decays of Z boson and Higgs boson.

\section{Probing prospects in future LHC }

The discovery potential at 14 TeV LHC are discussed in this section. Although the $\chi^0_{2,3}\chi^\pm_1 $ production rate will increase at 14 TeV LHC, the background will enlarge too. We must simulate the backgrounds as well as the signals. The irreducible background includes diboson, triboson and $t\bar{t}W/Z$ production, among which the diboson production highly dominates.
The reducible background includes single and pair production of top quarks, $WW$ and single $W$ or $Z$ boson processes produced in association with jets or photons, among which the $t\bar{t}$ production highly dominates. In our paper we only simulate the mainly backgrounds: the diboson backgrounds and the $t\bar{t}$  backgrounds. We use \textsf{MadGraph} simulate our backgrounds and scale the cross sections to the next leading order \cite{NLO}. To make our backgrounds more accurate, we first simulate the backgrounds at 8TeV. After comparing our simulated backgrounds and the backgrounds derived from experimental results, we get the scale factors in each signal region. Then we multiply the corresponding scale factors on our simulated 14TeV backgrounds and then we take these scaled 14TeV backgrounds as our backgrounds. Although at 14 TeV LHC the scaled factors might be changed a little, it still offset some deviation between our simulations and the experimental background estimation.

In Table 1 lists the number of background events at 14 TeV LHC 300 fb$^{-1}$. In the $Z$-enriched region, $WZ$ production dominates the background, whereas in the $Z$-depleted region, the $t\bar{t}$ has a comparative contribution.

\begin{table}[th] \centering\caption{The number of background events at 14 TeV LHC 300 fb$^{-1}$. \label{tab1}}
\begin{tabular}{|c|c|c|c|c|c|c|c|}
\hline  Background                 & SRnoZa         & SRnoZb      & SRnoZc   & SRZa  & SRZb & SRZc \\
\hline  $ZZ$                         &410 	&59 	&10 	&280 	&39 	&12  \\
\hline  $ZW^\pm$                   &1391 	&595 	&71 	&6850 	&661 	&189  \\
\hline  $t\bar{t}$                &1715 	&401 	&62 	&272 	&178 	&19  \\
\hline  Total                 &3516 	&1055 	&143 	&7402 	&878 	&220  \\
\hline
\end{tabular}
\label{tab2}
\end{table}

For the signal, the cross section are calculated using \textsf{Prospino2.1} \cite{prospino}. We implement the same cuts on the signal and backgrounds. The following formulas are adopted to calculate the significance
\begin{eqnarray}
  \text{Significance} = \frac{S}{\sqrt{B+(0.1 B)^2}}
\end{eqnarray}
where $S$ is the number of signal events and $B$ is the total number of background events. We also considered 10\% sys. error in the estimation.

\begin{figure}[htbp]
\includegraphics[width=4.0in,height=3in]{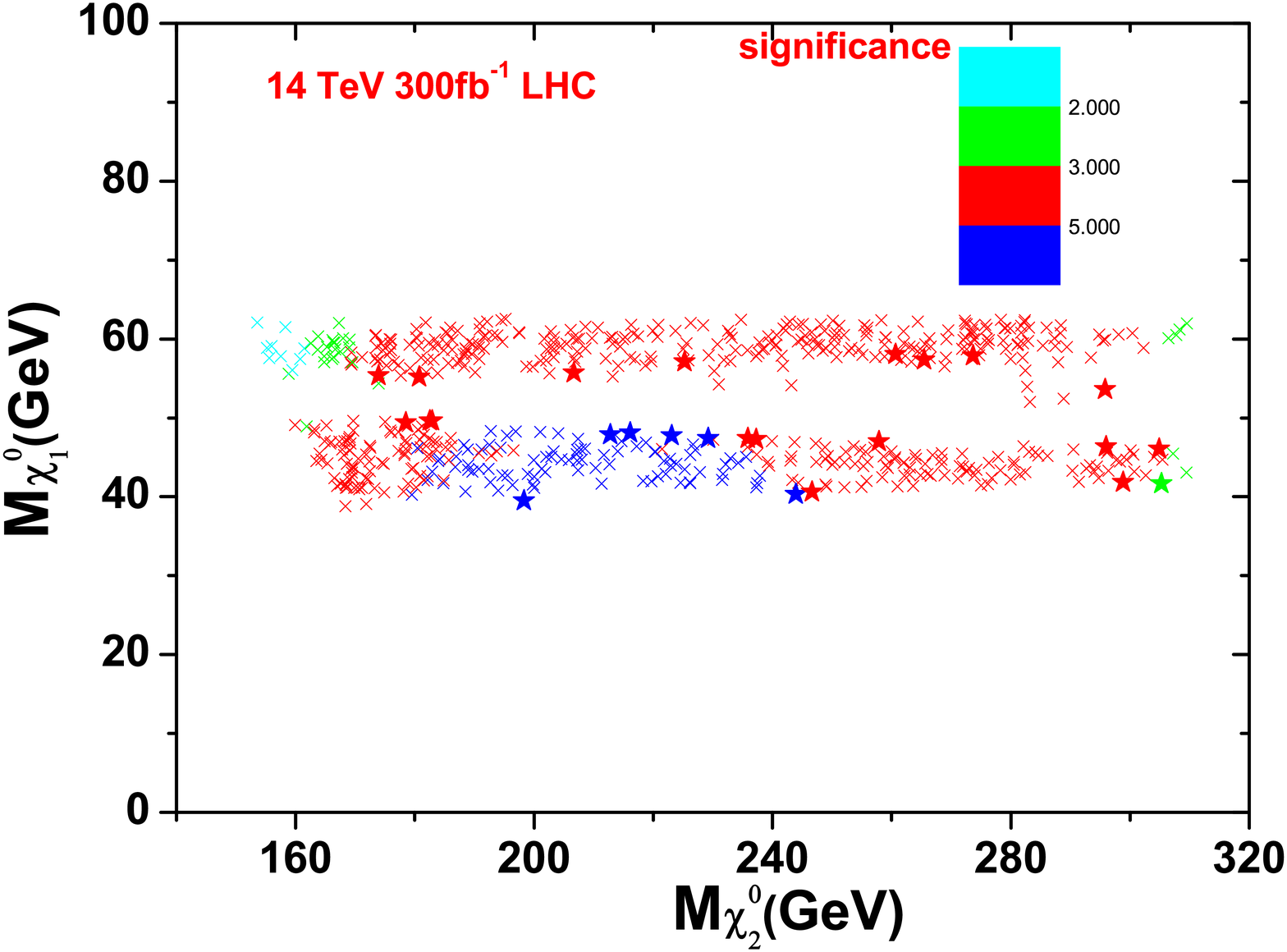}
\caption{The discovery potential at 14 TeV LHC with 300 fb$^{-1}$. The pentagrams on the plot satisfying the 2$\sigma$ range of PLANCK. }
\label{fig5}
\end{figure}

We present the final results in Fig. \ref{fig5}. It shows that the region with $40~\text{GeV} \lesssim m_{\tilde{\chi}^\pm_1} \lesssim 60~\text{GeV}$ and $160~\text{GeV} \lesssim m_{\tilde{\chi}^0_{2,3}} \lesssim 300~\text{GeV}$ can be covered at $3\sigma$ level. Some parameter space can reach the $5\sigma$ discovery level.  We note here the points satisfying the 2$\sigma$ range of PLANCK would be easily covered at $2\sigma$ level.
A tiny part of the parameter space is under $2\sigma$ because it locates at the region where the kinematics similar to the $WZ$ background. However, if the luminosity increases, this region would be more squeezed.

Finally, we stress that our analysis is performed in the framework of MSSM.
In some extensions of the MSSM, such as the next-to-minimal supersymmetric standard model (NMSSM) which
seems to be more favored by the LHC Higgs data \cite{nmssm-1},
the neutralino LSP may have a significant singlino component and thus can be very light \cite{nmssm-2}.
Then the $\tilde{\chi}^\pm_1 \tilde{\chi}^0_{2,3}$ production may have different signatures.

\section{Conclusion}

Motivated by the naturalness, we study a simplified MSSM scenario where only the bino-like LSP
and higgsino-like NLSP are light. We first scan the parameter space of this scenario,
considering the constraints from the Higgs mass, flavor physics, electroweak precision
measurements and dark matter experiments. Then in the allowed parameter space, we perform
a Monte Carlo simulation for the $\tilde{\chi}^\pm_1 \tilde{\chi}^0_{2,3}$ production followed by
$\tilde{\chi}^\pm_1 \to W^\pm \tilde{\chi}^0_1$ and $\tilde{\chi}^0_{2,3} \to Z\tilde{\chi}^0_1$.
By examining the presently available trilepton bounds on the wino-like chargino/neutralino,
we find that only a narrow region $40\,\rm{GeV} \lesssim m_{\tilde{\chi}^0_1} \lesssim 50\,\rm{GeV}$ and $160\,\rm{GeV} \lesssim m_{\tilde{\chi}^0_{2,3}} \lesssim 170\,\rm {GeV}$ on the plane of
$m_{\tilde{\chi}^0_1}-m_{\tilde{\chi}^0_{2,3}}$ can be excluded. Finally, we explore the potential
of trilepton signature in probing such a scenario at 14 TeV LHC and find that the region
with $40\,\rm{GeV} \lesssim m_{\tilde{\chi}^0_1} \lesssim 60\,\rm {GeV}$ and
$160 \rm {GeV}\,\lesssim m_{\tilde{\chi}^0_{2,3}} \lesssim 300\,\rm{GeV}$ can be covered
at $3\sigma$ level with luminosity ${\cal L}=300$ fb$^{-1}$.

\section*{Acknowledgement}
I would like to thank Junjie Cao, Jie Ren,  Lei Wu, Jin Min Yang and Yang Zhang for helpful
discussions and valuable comments on the manuscript. I acknowledge the Korea Ministry of Education,
Science and Technology (MEST) for the support of the Young Scientist Training Program at the Asia
Pacific Center for Theoretical Physics (APCTP).

\hbox to \hsize{\hss}

\end{document}